\documentclass[twocolumn]{aastex63}

\shorttitle{Onset of Fast Reconnection in the Solar Transition Region}
\shortauthors{Guo et al.}

\usepackage{natbib}
\accepted{26 August 2020}
\submitjournal{ApJ}

\newcommand{\longacknowledgment}{We gratefully acknowledge support by NASA contract NNG09FA40C (IRIS).
The simulations have been run on the Pleiades cluster through the
computing project s1061 from the High End Computing (HEC) division of
NASA. Data are courtesy of IRIS, SDO/AIA and SDO/HMI. IRIS is a NASA
small explorer mission developed and operated by LMSAL with mission
operations executed at NASA Ames Research Center and major
contributions to downlink communications funded by ESA and the
Norwegian Space Centre.}

\begin{document}

\title{Observations and modeling of the onset of fast reconnection in
  the solar transition region}

\correspondingauthor{Bart De Pontieu}
\email{bdp@lmsal.com}


\author{L.-J. Guo}
\affil{Lockheed Martin Solar \& Astrophysics Laboratory,
3251 Hanover St, Palo Alto, CA 94304, USA}
\affil{Bay Area Environmental Research Institute,
NASA Research Park, Moffett Field, CA 94035, USA.}

\author{Bart De Pontieu}
\affil{Lockheed Martin Solar \& Astrophysics Laboratory,
3251 Hanover St, Palo Alto, CA 94304, USA}
\affil{Rosseland Center for Solar Physics, University of Oslo, P.O. Box 1029 Blindern, N-0315 Oslo, Norway}
\affil{Institute of Theoretical Astrophysics, University of Oslo,
P.O. Box 1029 Blindern, N-0315 Oslo, Norway}

\author{Y.-M. Huang}
\affil{Center for Heliophysics, Princeton Plasma Physics Laboratory,
  Princeton University, Princeton, NJ 08544}
\affil{Max-Planck-Princeton Center for Plasma Physics}

\author{H. Peter}
\affil{Max Planck Institute for Solar System Research,
  Justus-von-Liebig-Weg 3, 37077 Göttingen, Germany}
\affil{Max-Planck-Princeton Center for Plasma Physics}

\author{A. Bhattacharjee}
\affil{Center for Heliophysics, Princeton Plasma Physics Laboratory,
  Princeton University, Princeton, NJ 08544}
\affil{Max-Planck-Princeton Center for Plasma Physics}

\begin{abstract}
  Magnetic reconnection is a fundamental plasma process that plays a
  critical role not only in energy release in the solar atmosphere,
  but also in fusion, astrophysical, and other space plasma
  environments. One of the challenges in explaining solar observations
  in which reconnection is thought to play a critical role is to
  account for the transition of the dynamics from a slow
  quasi-continuous phase to a fast and impulsive energetic burst of
  much shorter duration. Despite the theoretical progress in
  identifying mechanisms that might lead to rapid onset, a lack of
  observations of this transition has left models poorly constrained.
  High-resolution spectroscopic observations from NASA’s
  Interface Region Imaging Spectrograph (IRIS) now reveal tell-tale
  signatures of the abrupt transition of reconnection from a slow
  phase to a fast, impulsive phase during UV bursts or explosive events in the
  Sun’s atmosphere. Our observations are consistent with numerical
  simulations of the plasmoid instability, and provide
  evidence for the onset of fast reconnection mediated by plasmoids
  and new opportunities for remote-sensing diagnostics of reconnection
  mechanisms on the Sun.
\end{abstract}


\section{Introduction} \label{sec:intro}
Magnetic reconnection is a process in which the magnetic topology is
changed and magnetic energy is released to heat plasma, drive Alfv\'enic
flows, induce Alfv\'en as well as magneto-acoustic waves and energize
particles. It is responsible for energetic phenomena such as
geomagnetic storms and aurora \citep{1,Sitnov2019}, solar flares and coronal mass
ejections \citep{2,Shibata2011,3,4,5,6}, X-ray flares in magnetars \citep{7}, and magnetic
interactions between neutron stars and their accretion disks \citep{8}. It
is also crucially important in laboratory plasma physics because it
triggers the sawtooth crashes that can potentially cause disruptive
instability in magnetically confined fusion plasmas \citep{9,Bhattacharjee2001,10}. A better
understanding of reconnection is thus a common challenge confronting
laboratory, space, and astrophysical plasma physicists.

\begin{figure*}
	\includegraphics[width=0.95\textwidth]{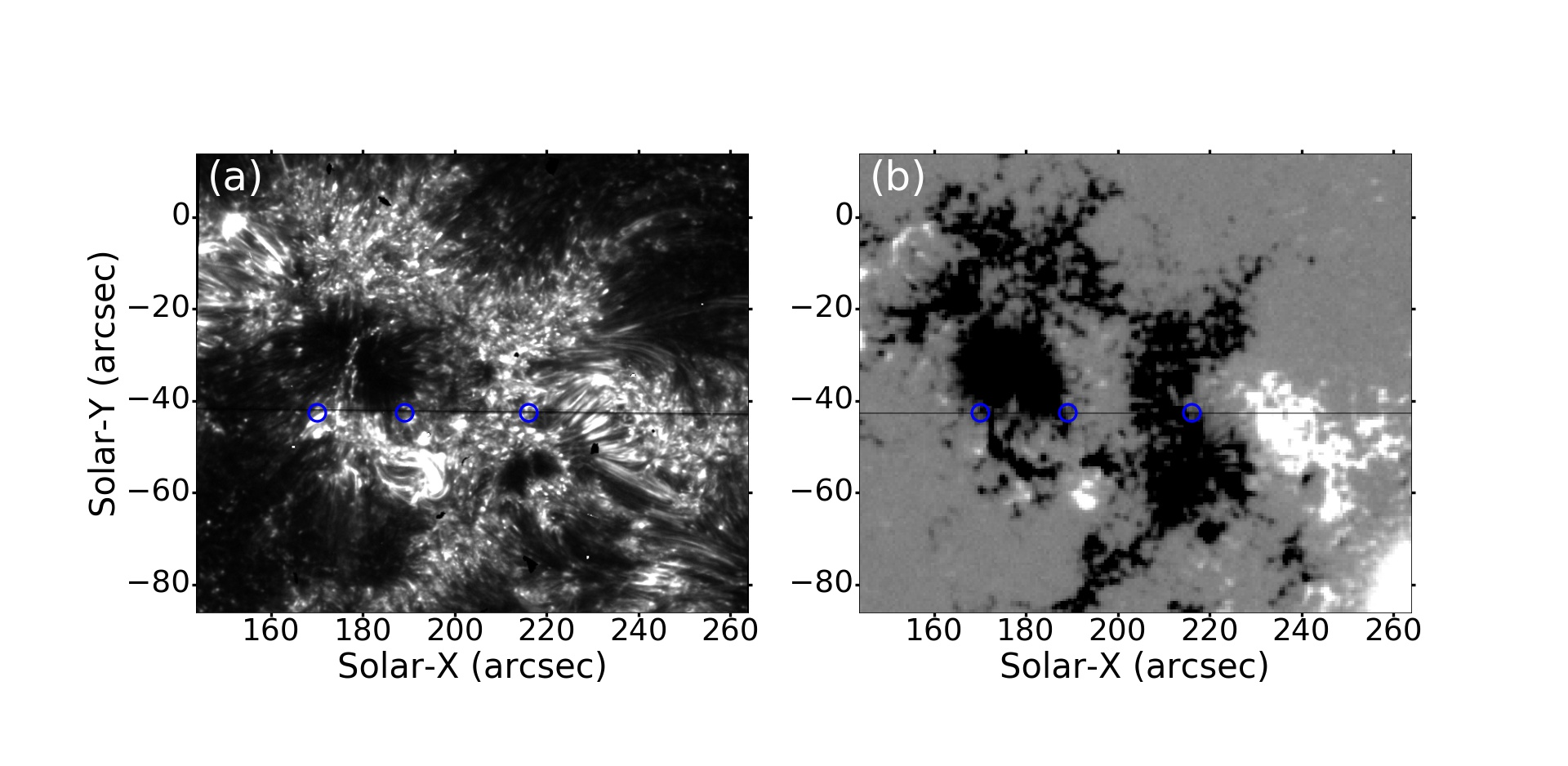}  
	\caption{\label{fig:overview} UV bursts occur in regions of strong magnetic
          activity. Active Region 12036 on 2014 May 05 13:22:21 UT:
          (a) IRIS 1400 slit jaw image in a wavelength range around
          1400\AA\ showing mainly 80,000 K plasma; (b) HMI
          line-of-sight magnetogram showing the distribution of
          magnetic field on the solar surface. The black horizontal
          line in (a) and (b) is the IRIS spectrometer slit, the blue
          ‘O’ signs mark locations of UV bursts.  When observed at sufficiently high spatial
          resolution and magnetic sensitivity (i.e. higher than
          displayed here), these bursts are often located where
          opposite magnetic polarities meet.
        }
\end{figure*}

Recently, observational and experimental works have provided new
insights into reconnection mechanisms. For example, NASA’s
Magnetospheric Multiscale Mission (MMS) has found direct evidence for
electron demagnetization and acceleration at reconnection sites in
Earth’s magnetosphere \citep{11}, while laboratory experiments have revealed
structures in the diffusion region that trigger fast reconnection \citep{12,
  57, 56}. Observations from the Cluster spacecraft in the Earth's
magnetotail \citep{Chen2008}
and more recently the Parker Solar Probe have found magnetic islands in the
inner solar wind \citep{53}. However, many questions remain about how the onset of fast reconnection occurs. For example, transition region explosive
events, rapid brightenings thought to be driven by reconnection and associated with strong flows visible in
spectral lines formed at temperatures of a few hundred thousand Kelvin
\citep{13, 14}, occur on timescales of minutes, much shorter than classic
reconnection mechanisms predict. Recent progress in reconnection
theory \citep{Shibata2001,15, 16} suggests that under certain circumstances fast
reconnection can be triggered by the so-called plasmoid instability,
when a thin current sheet breaks up into secondary current sheets
and plasmoids (or magnetic islands). Recent theoretical studies \citep[e.g.,][]{
Pucci2013,28, 29, 30} suggest that the thinning of the current sheet and growth
of various tearing modes proceed simultaneously before fast
reconnection is triggered. Nevertheless, these predictions for the
onset of the fast reconnection, namely a transition from a slow
quasi-continuous phase to fast, plasmoid-mediated, reconnection, have
not yet found definitive support from observations, despite
observations of the presence of plasmoid-like features \citep{17, 18,
  19, Rouppe-van-der-Voort2017}. In addition, alternative theories
have suggested that strong turbulent motions in solar microflares
lead to the onset of fast reconnection \citep{Chitta2020}. 

UV bursts, sudden and compact brightenings in UV light,
provide an opportunity to study magnetic reconnection on the Sun
\citep{54}. A variety of phenomena such as transition region explosive
events and IRIS bombs are considered to be UV bursts: we shall refer
to all these events as UV bursts in this paper. These bursts
are often characterized by broad spectral line profiles
and the more violent ones show a strong enhancement of emission
\citep{20}. They are thought to be driven by magnetic reconnection \citep{20, 21,
22}, because of their characteristic spectral profiles that indicate
strong bidirectional flows, which typically occur where magnetic flux
concentrations of opposite polarity meet \citep{55}. Other explanations for
explosive events exist as well \citep{23, 24}, but the majority of explosive
events are still considered to be driven by reconnection. As
mentioned, high-resolution observations of UV bursts have provided
tantalizing indications for the presence of
plasmoids \citep{Rouppe-van-der-Voort2017}, but they have not shown
the transition from a slow to fast phase. In this paper we present
observational evidence for such a transition, and compare it with
numerical simulations of magnetic reconnection, transitioning from a
slow phase to a plasmoid-mediated phase in which reconnection occurs
much faster.

In \S~\ref{sec:obs} we describe the observations, while we provide
details of the numerical simulations in \S~\ref{sec:sim}. Our results
are described in \S~\ref{sec:res} and we finish with a discussion and
conclusions in \S~\ref{sec:con}.

\section{Observations}\label{sec:obs}

IRIS is a NASA small explorer that contains a high-resolution 
spectrograph and slit-jaw imager, both sensitive to near- and far-UV light.
The instrument has spectral passbands that cover emission lines
and continua that originate at temperatures covering the solar
atmosphere from the photosphere to the corona \citep{25}. In this paper, we
concentrate on the \ion{Si}{4} 1403\AA\ line originating in the transition
region between the chromosphere and corona at about 80,000K. The
spatial resolution of IRIS is 0.33\arcsec, while the spectral sampling
corresponds to a velocity sampling of around 2.7 km/s. We also use the
slit-jaw images of the 1400\AA\ passband from the IRIS slit-jaw imager
which are dominated by the \ion{Si}{4} lines at 1394\AA\ and 1403\AA\, as well as
the far ultraviolet continuum \citep{25}. We analyzed four IRIS data
sets (so-called sit-and-stare rasters) to search for UV bursts sharing similar spectral properties: 13-Aug-2013
at 13:35 UTC, 4-Feb-2014 at 00:29 UTC, 15-Apr-2014 09:59 UTC, and
4-May-2014 12:09 UTC. These datasets have cadences short enough (3s
for all, except for the latter, which has 5s cadence) to
observe fast-evolving UV bursts. All were targeted on active
regions. 

\begin{figure*}
  \includegraphics[width=0.5\textwidth]{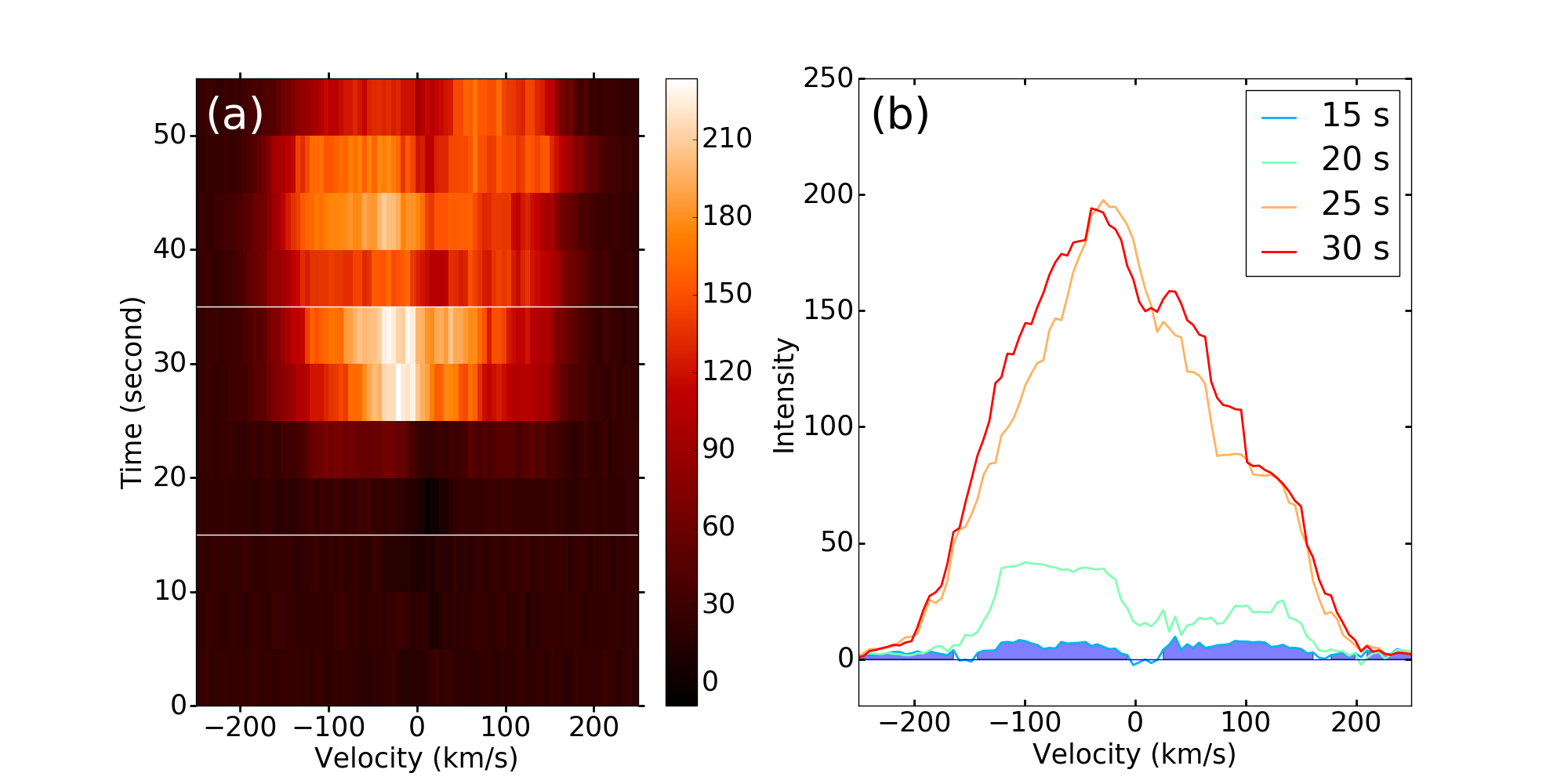}
  \includegraphics[width=0.5\textwidth]{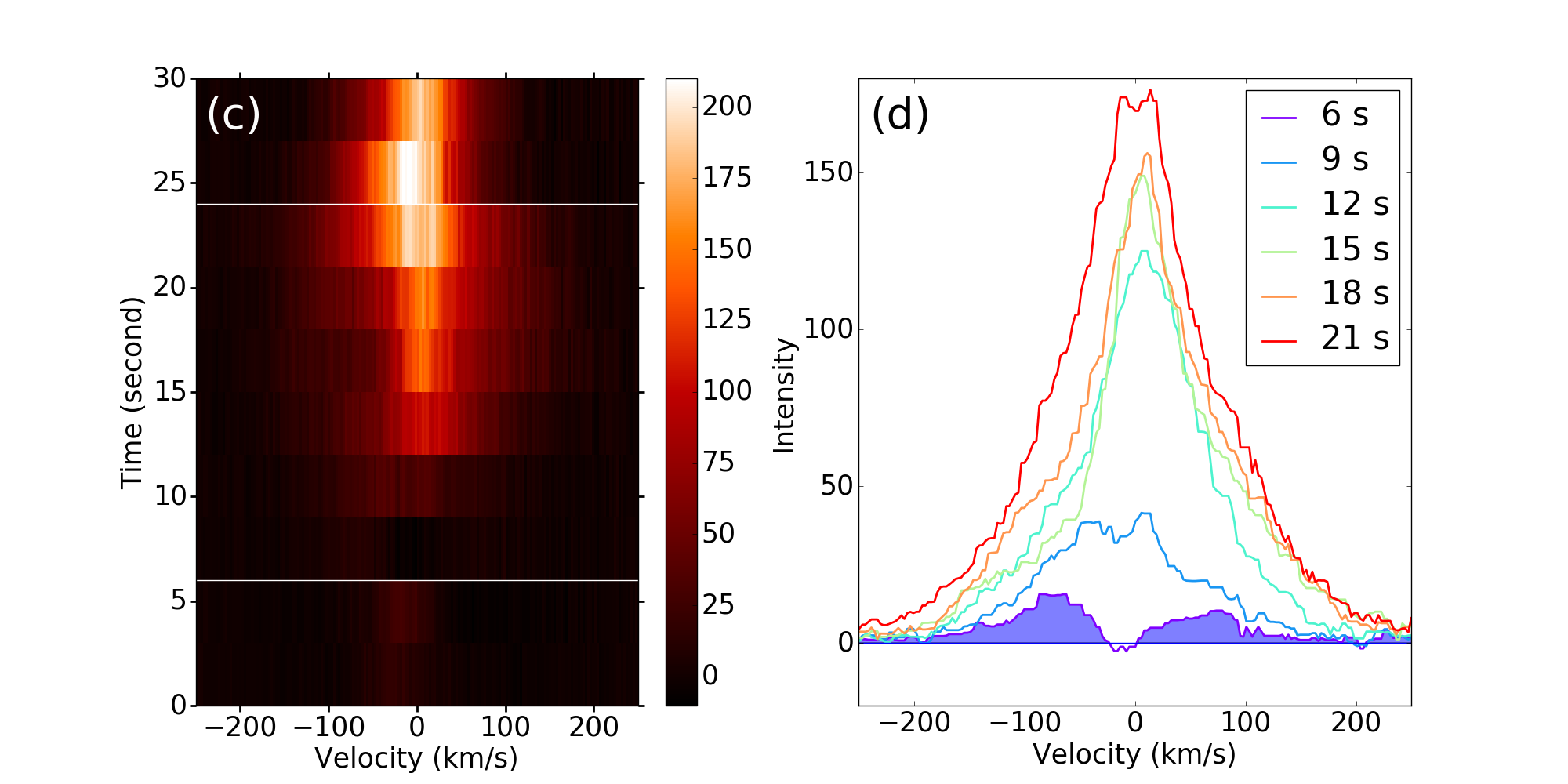}
  \\
  \includegraphics[width=0.5\textwidth]{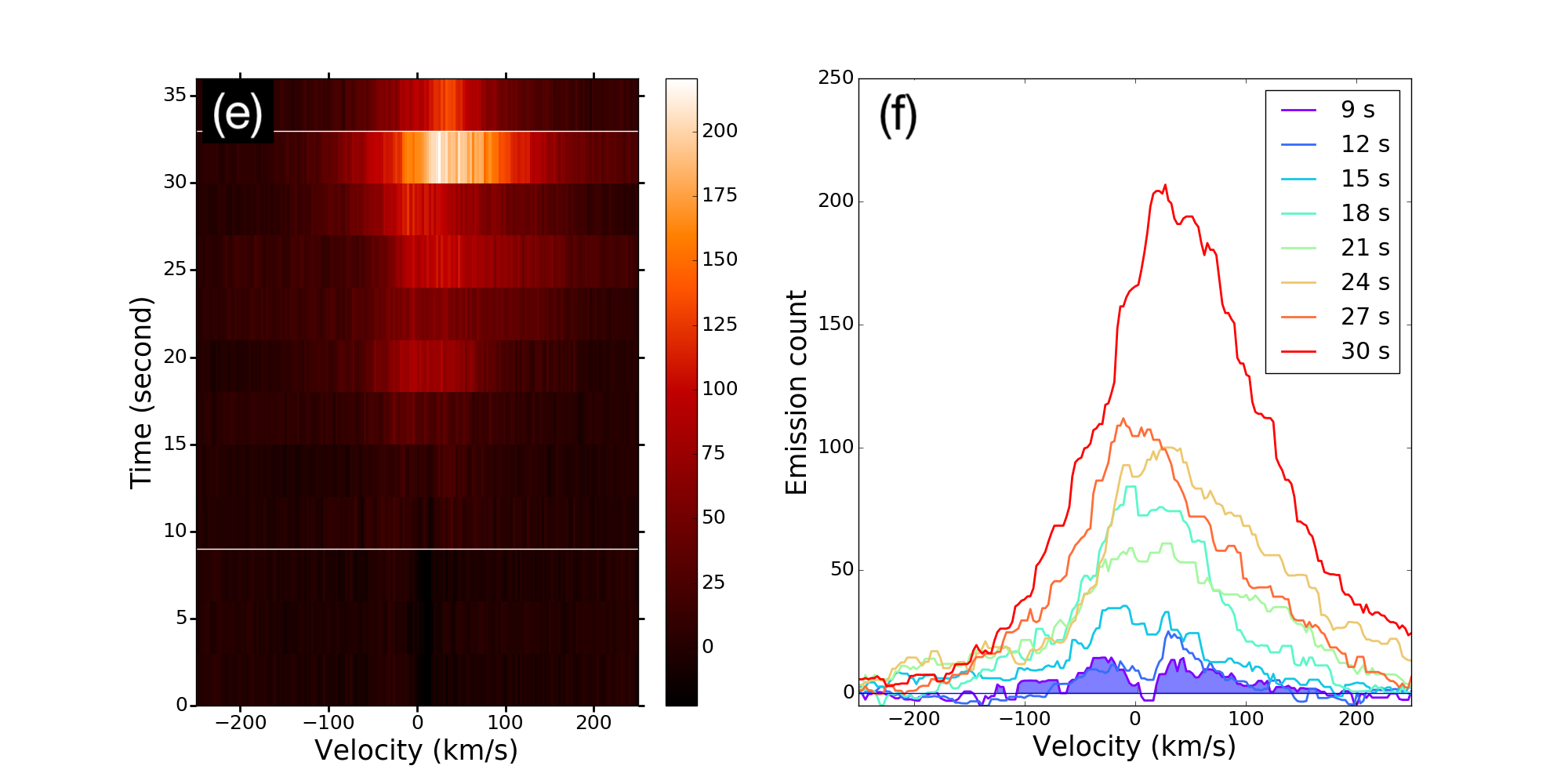}
  \includegraphics[width=0.5\textwidth]{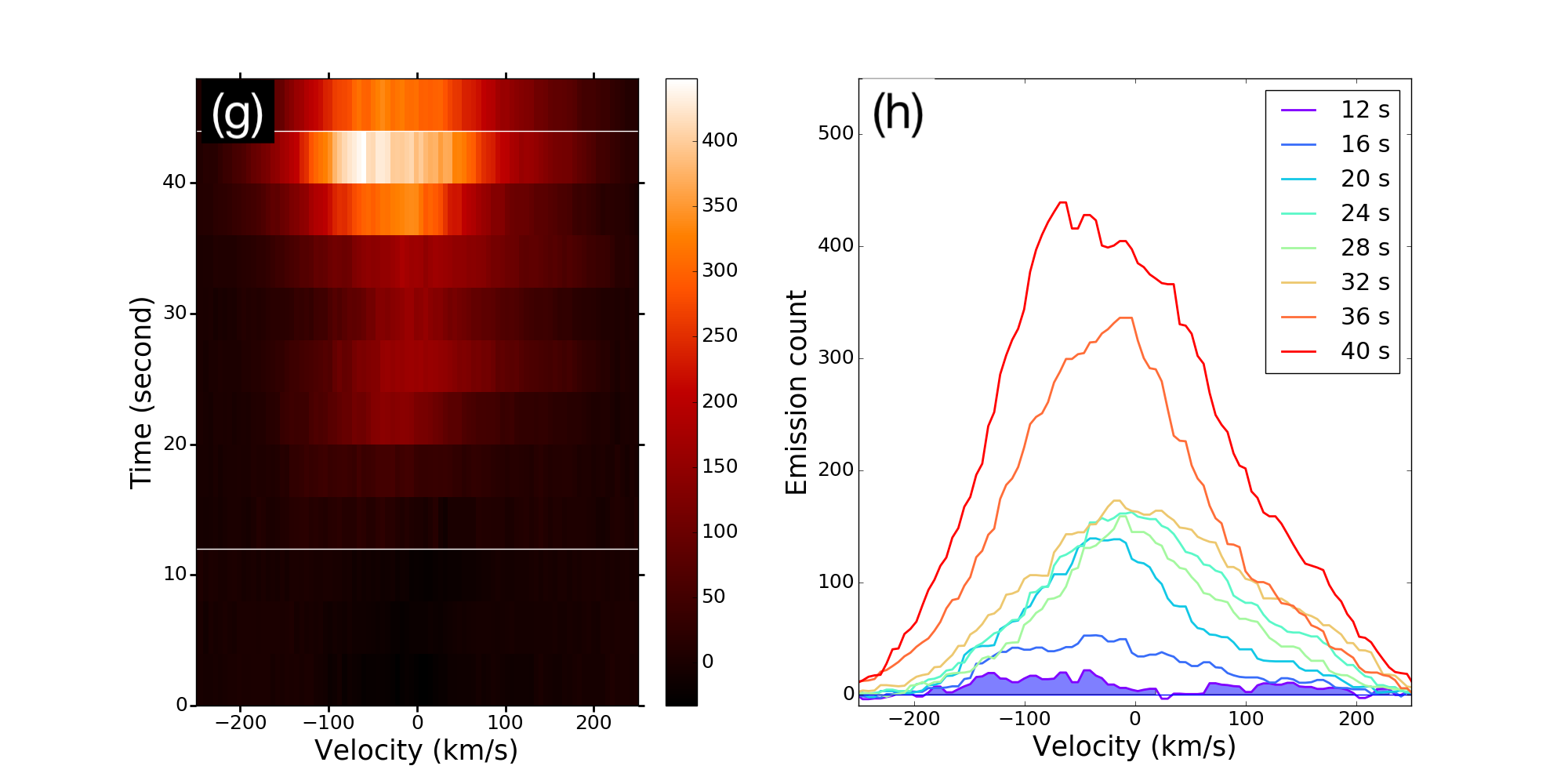}

	\caption{\label{fig:obs1} IRIS observations of UV bursts indicate the onset of fast
          reconnection. Panel (a): time evolution of \ion{Si}{4} spectra at
          the beginning of the event. Panel (b): spectral line
          profiles for selected times in the time range highlighted by white horizontal
          lines in panel (a). Panels c, d, e, f, g, and h show similar
          evolution for three other events. All four events show transitions
          from low-intensity, double-peak shaped line profiles to
          high-intensity, triangle shaped line profiles. In panels b,
          d, f, and h the line profile at the first time step is shaded
          with blue to accentuate its double-peak shape.
        }
\end{figure*}

We also use a magnetogram obtained by the Helioseismic and Magnetic
Imager \citep[HMI,]{31} onboard the Solar Dynamics Observatory \citep[SDO,][]{32}
along with an IRIS slit-jaw image to show an active region with
several UV bursts in Figure~\ref{fig:overview},
with sample spectra shown in Fig.~\ref{fig:obs1}. Co-alignment of the HMI and IRIS data was done by cross-correlation
between the 1600\AA\ continuum obtained by the Atmospheric Imaging
Assembly \citep[AIA]{33} onboard SDO and IRIS 2796\AA\ slit-jaw images that
show the chromosphere in the Mg II k line. We first
perform co-alignment using the coordinates provided in the headers of
the data sets and then use a cross correlation method to determine the
pointing offset between the IRIS telescope and the AIA telescope. The
HMI and AIA data are both full-disk and easily co-aligned.

\section{Simulations}\label{sec:sim}

\begin{figure*}
  \includegraphics[width=\textwidth]{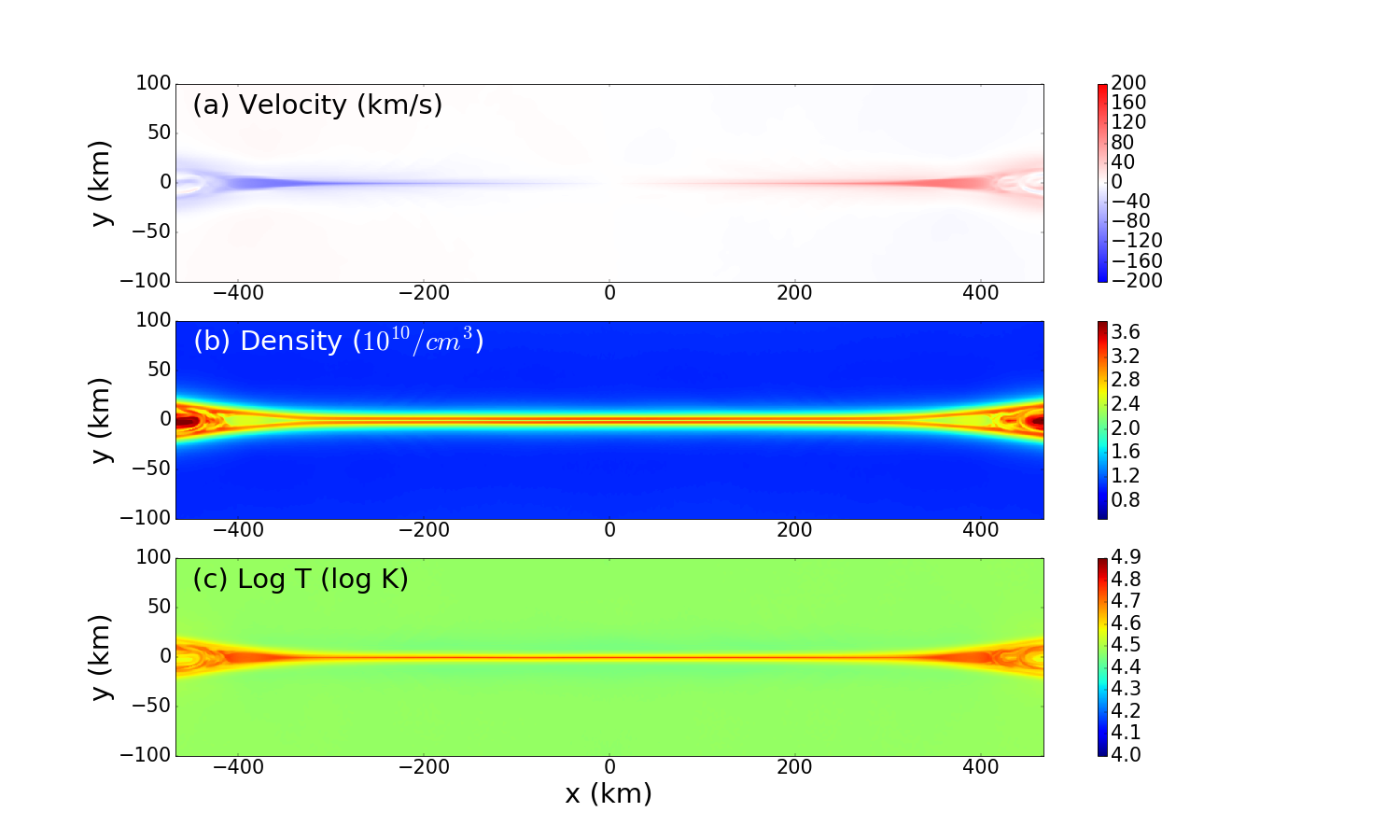}\\
  \includegraphics[width=\textwidth]{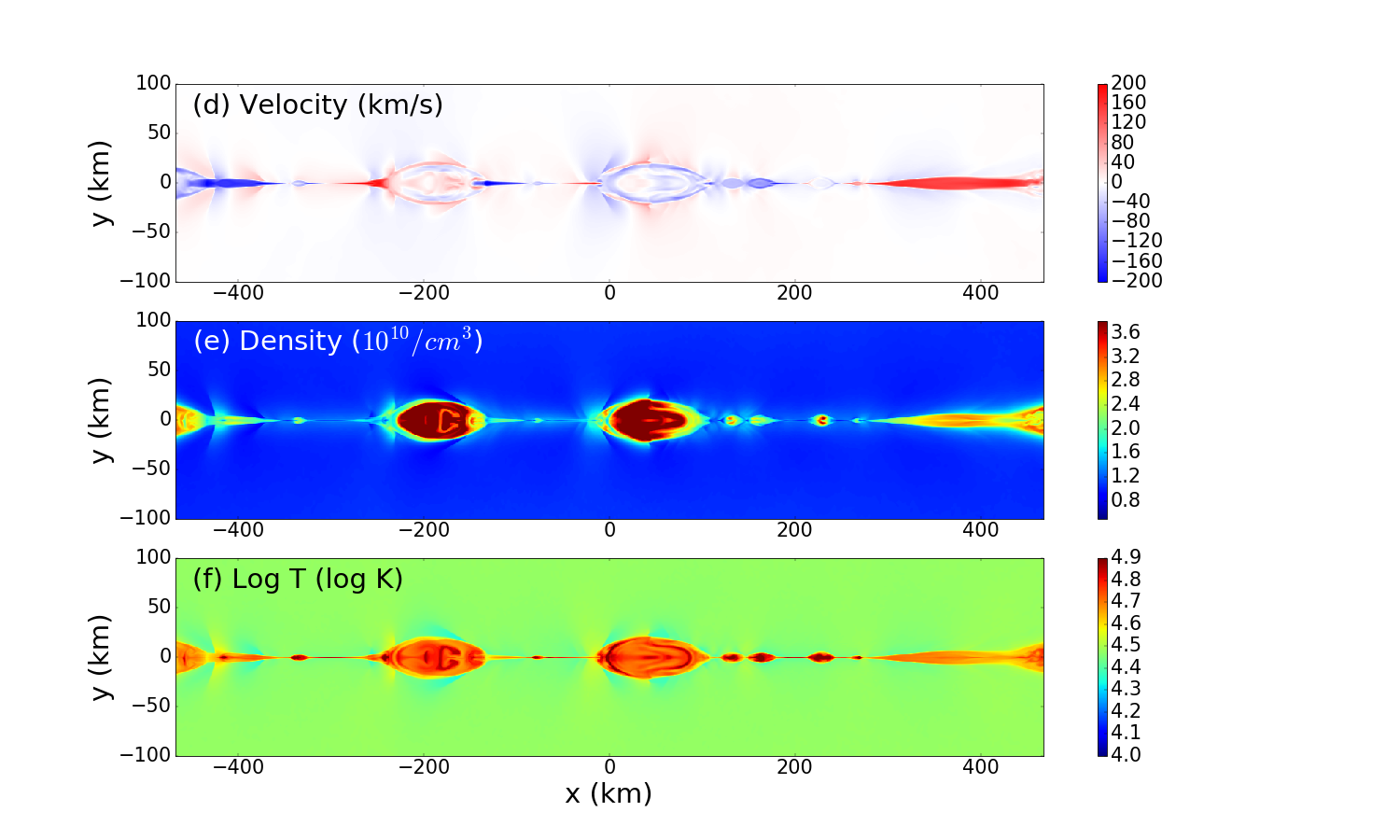}
	\caption{\label{fig:sim} A numerical simulation shows the
          transition from slow reconnection (left) to fast
          reconnection mediated by plasmoids (right). Panels (a), (b)
          and (c): velocity, density and temperature during the slow
          phase. Panels (d)-(f) show the same during the fast
          phase. Please note that the aspect ratio is enhancing the
          distance in the y-direction. The domain shown here covers
          about 1 Mm by 0.2 Mm, while the entire simulation domain
          covers 2 Mm by 2 Mm. The time between these two snapshots of
          the slow (t=4s in Fig.~\ref{fig:sim2}) and the fast phase (t=10s in
          Fig.~\ref{fig:sim2}) is only 6 seconds. }
\end{figure*}

The simulations in this letter solve single-fluid, resistive MHD
equations with anisotropic heat conduction and radiative cooling,
using a state-of-the-art simulation code \citep{30}. Similar to
\citet{26}, we use a Harris sheet
equilibrium as the initial configuration, which is given by $B_x(y) =
B_0 \tanh(y/\delta)$ \citep{34}
where $\delta=10 km$  is the half width of the current sheet; we adopt a Lundquist
number of $10^5$, which is defined by $S = L v_A / \eta $ with L the spatial
scale, $v_A$ the Alfv\'en speed and $\eta$ the resistivity.
We adopt the radiative loss estimated in \citet{36}. Assuming the simulation
domain is in thermal equilibrium initially,
background heating is added to balance the radiative loss at t=0
s. The resistive MHD equations are solved in a
two-dimensional domain 2~Mm $\times$ 2~Mm on a grid of 2064 $\times$ 2048 grid
points. In Figure \ref{fig:sim} we show a subdomain enclosing
the current sheet ($<1$ Mm) instead of the whole simulation domain. The
$x$ direction is along the long axis of the current sheet and the $y$
direction perpendicular to the current sheet. Viscosity is not
included in this qualitative study. Initially, at the asymptotic
region, the Alfv\'en speed is 200 km/s, the number density $10^{10}$ cm$^{-3}$, the
temperature $3\, 10^4$ K and the plasma beta = 0.1. We did not include
gravity and a stratified atmosphere in the current study. The fact
that the spatial extent of a UV burst, and thus our
computational domain, is smaller than the pressure scale height in the
transition region at around 100,000 K justifies this
assumption. Gravity and stratification would cause a density gradient
along the outflow direction and possibly decrease the speed of
reconnection outflows toward the Sun and slightly alter the red wings
of line profiles. Nevertheless, we do not expect that including these
two terms will qualitatively change our results.

Note that the simulation is conducted over a relatively narrow range
of plasma parameters, which are chosen specifically to focus on the
observed events in the temperature range of \ion{Si}{4} formation. The
initial temperature is chosen to be 30,000 K so that the current sheet is
at the lower range of the \ion{Si}{4} contribution function (weaker
emission)  in the slow phase and at the peak of the \ion{Si}{4} contribution
function (stronger emission) in the fast phase. If we use lower
initial temperatures (e.g. 10,000 K), the slow phase is less visible in
\ion{Si}{4} but the emission in the fast phase will still be strong. It
certainly is possible to imagine other scenarios in which the slow
phase is too cool to be visible in the \ion{Si}{4} passband, the setup of
our simulation in this paper is focused on UV bursts observed
in \ion{Si}{4} with both phases visible.

\section{Results}\label{sec:res}
\subsection{Observations of onset of fast reconnection}
A large number of UV bursts are observed
in the four datasets. We use a feature-detection code to select line
profiles with large line width and intensity, and filter out
asymmetric line profiles, which are likely caused by events that are
spread over several resolving elements of IRIS and would complicate
analysis. Another advantage of selecting symmetric line profiles is
that they are compatible with the choice of the line-of-sight
direction along the outflows and our assumption of the numerical
domain being enclosed within one IRIS pixel. Among all the events that
satisfy the above criteria, we pick events with visible sharp onset,
judging from changes of intensity and line width. Following the above
selection criteria, we have found eight events in our datasets, of which
we show four in Fig.~\ref{fig:obs1}.

\begin{figure*}
  \includegraphics[width=\textwidth]{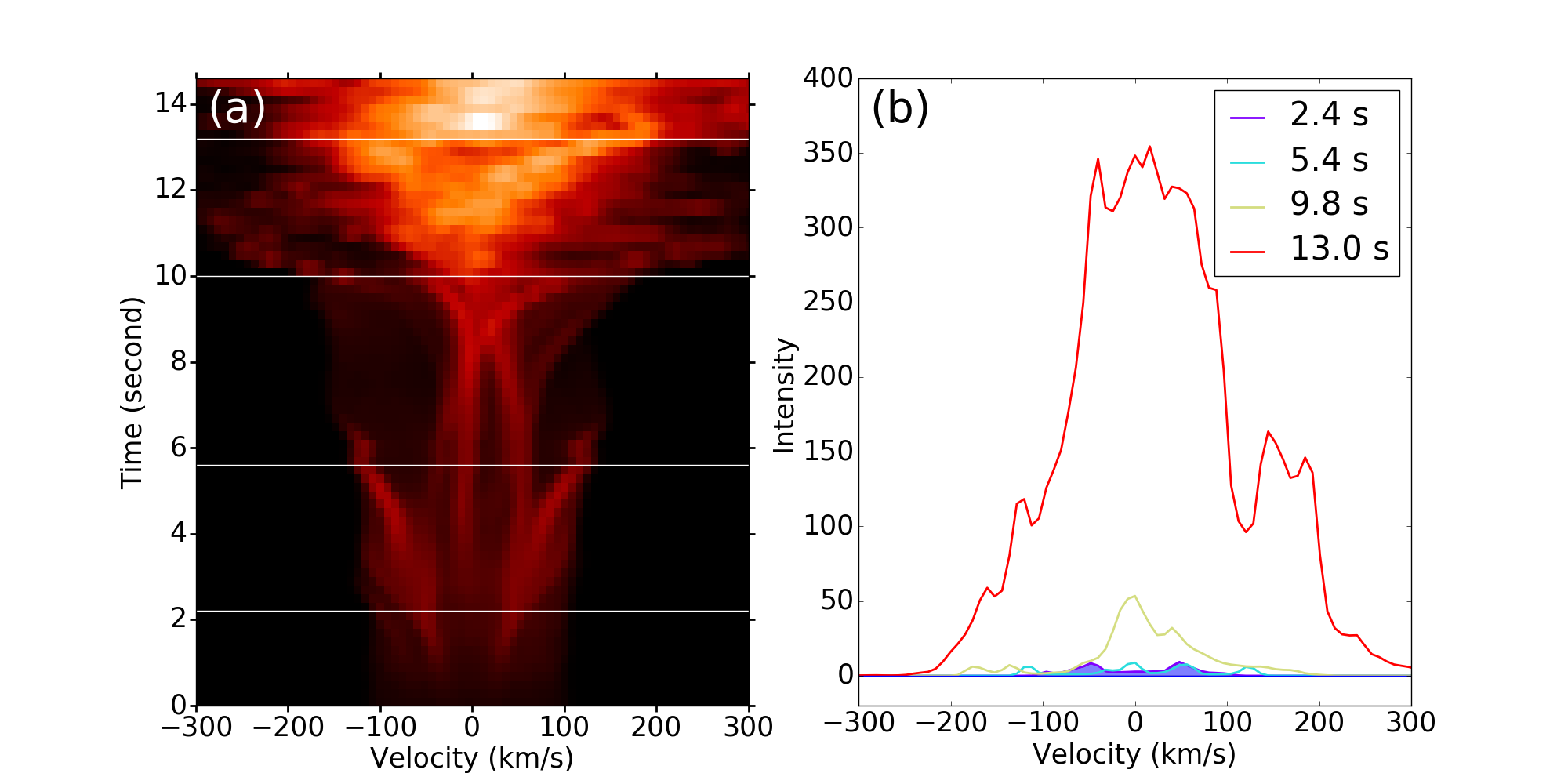}

	\caption{\label{fig:sim2} Synthesized \ion{Si}{4} spectra from the
          model agree well with IRIS observations. Panel (a) shows the
          time evolution of \ion{Si}{4} spectra synthesized from the
          reconnection model. Panel (b) shows the line profiles of the
          time steps marked by white horizontal lines in panel
          (a). These synthetic line profiles compare well in a
          qualitative sense with the observations shown in Fig.~\ref{fig:obs1}, even
          though panel (a) is richer in structure here in the
          simulation than in the observations (Fig.~\ref{fig:obs1}a,c) which is
          because of the higher resolution of the numerical model.
        }
\end{figure*}

For events from data sets with short exposure times, line profiles can
be noisy due to a low signal to noise ratio. In this case, we run a
four-point averaging in the spatial and spectral dimensions to reduce
the noise level in the data. To enhance the visibility of any signals
immediately preceding the bright UV bursts, we 
subtract a line profile
that is averaged over 10 time steps before the onset of the event to
remove emission contributions from unrelated events that are either
along the line of sight, or introduced by scattered light from
neighboring pixels.

A few examples of UV bursts in the \ion{Si}{4} 1400\AA\ and their
locations overplotted on the photospheric magnetogram are shown in
Figure~\ref{fig:overview}. These events hold the key in revealing the dynamics of
reconnection sites and because transition region emission typically
forms in a small volume, line-of-sight superposition, a common issue
with observations in optically thin lines in the solar atmosphere, is
much less of a challenge for such emissions. Here we report on IRIS
observations of UV bursts in the \ion{Si}{4} 1400\AA\ line showing
rapidly evolving spectra that indicate the onset of fast
reconnection. Four examples are shown in Figure~\ref{fig:obs1}. Panel (a) shows the
time evolution of the \ion{Si}{4} spectra observed during the first event. A
sharp increase of intensity and width of spectra is visible at t=25s,
which indicates the beginning of the impulsive phase of
reconnection. Most interestingly, weak emission lasting two time steps
is observed before t=25s, revealing activity preceding the onset of
the impulsive phase. The intensity of line profiles at t=15
and t=20 is much weaker than line profiles at t=25 and t=30,
suggesting that before the impulsive phase, heating is weaker than
during the impulsive phase. Panel (b) of Figure~\ref{fig:obs1} shows line profiles
from t=15s to t=30s, with their positions marked by white lines in
panel (a), showing more details of the evolution of line profiles. At
t=15, the observed line profiles exhibit double peaks (see the area
shaded in blue), which can be produced by the bi-directional outflows
expected from single-X-point reconnection \citep{22}. The low
intensity at t=15s indicates weak activity, suggesting the
single-X-point reconnection proceeds at a slow rate. In the impulsive
phase, the most commonly observed line profiles are triangular line
profiles with wide wings and a strong central component at the rest
wavelength of \ion{Si}{4}, which we suggest is caused by fast reconnection
mediated by plasmoids \citep{26}. The impact of plasmoids on the observed
spectra has also been discussed by \citet{27,Rouppe-van-der-Voort2017}. Panel (c)
and panel (d) show the evolving line profiles for the second event,
whose line profiles undergo evolution similar to the first event but
exhibit  more temporal snapshots during the lifespan of the event. We
examined four different IRIS data sets of active regions and found
another six cases showing similar properties. Two of those cases are
shown in panels (e), (f), (g), and (h).

\begin{figure*}
  \includegraphics[width=\textwidth]{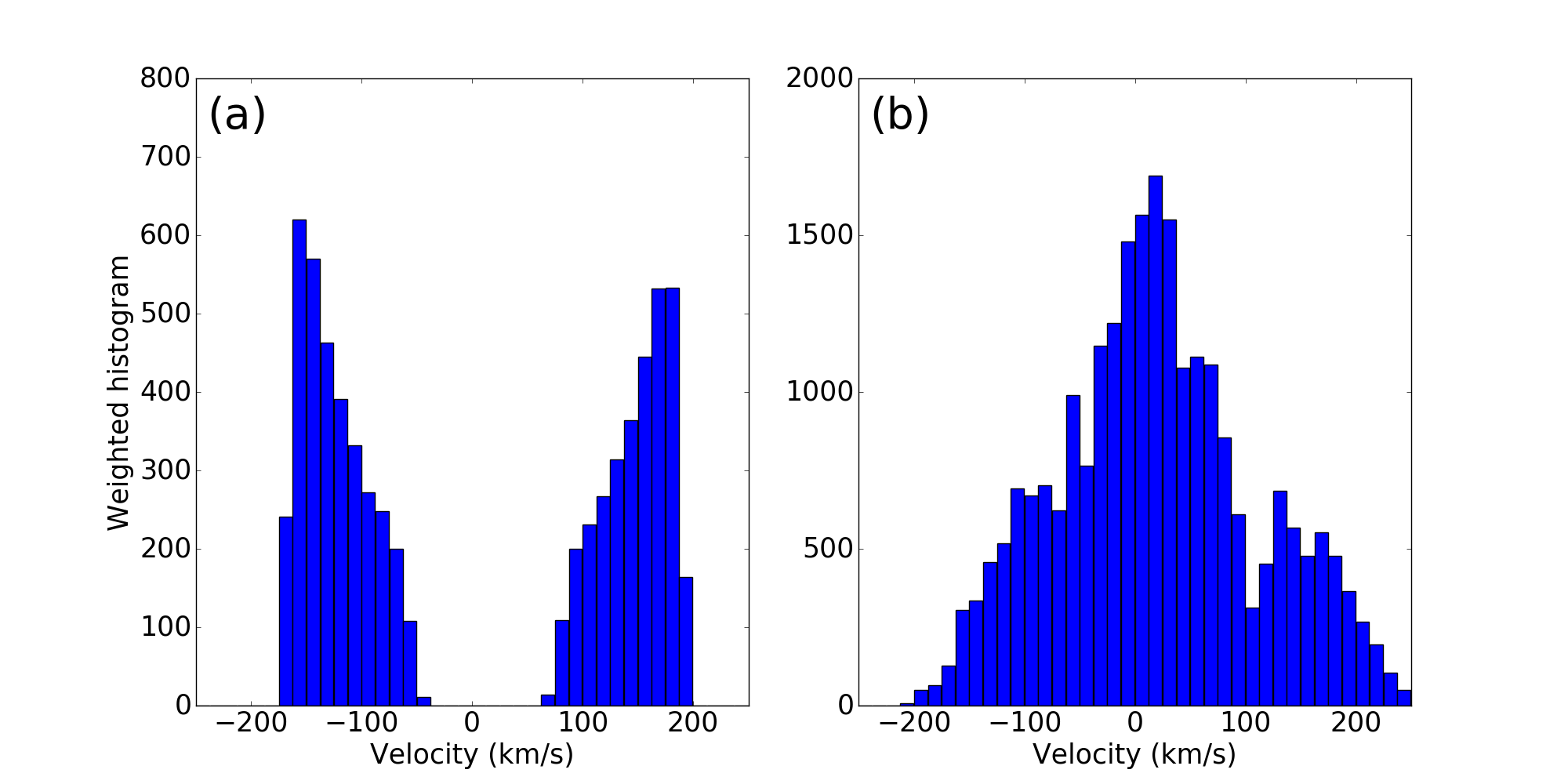}

	\caption{\label{fig:sim3} Weighted histogram of the outflow
          velocity (weighted by the square of the density)
          in the current sheet during the slow phase (a) and the fast
          phase (b).          
        }
\end{figure*}

\subsection{Numerical modeling of reconnection in transition region}
In order to better interpret observations of UV bursts, we
simulate a thinning current sheet with plasma parameters as close to
that of the transition region as presently enabled by a
state-of-the-art simulation code \citep{30}. In this simulation, the current
sheet gradually thins down as multiple tearing modes slowly grow
within it. When the islands produced by the most dominant mode disrupt
the formation of the current sheet, the reconnection dynamics make a
transition from the slowly evolving stage to rapid onset \citep{28, 29, 30, 51,
52}. Panels (a-c) of Figure~\ref{fig:sim} show the velocity, density and
temperature of the current sheet during the slow phase, when the
current sheet is dominated by an Alfv\'enic, bi-directional jet. The
bi-directional jet weakens when plasmoids start to appear in the
current sheet. As the plasmoids grow and coalesce, a substantial
amount of plasma is trapped in the plasmoids (see the density in panel
(e) of Fig.~\ref{fig:sim} during the fast phase), some of which moves at slow
velocities, compared to plasma in the secondary current sheet that
mostly moves at Alfv\'enic speeds (see the velocity in panel (d) during
the fast phase). As a result of the cascade of plasmoid formation, the
velocity distribution in the current sheet assumes a wide range of
values as well. With the velocity, density and temperature
distribution from the simulation, we are able to calculate the \ion{Si}{4}
spectra of the simulated current sheet. The synthetic \ion{Si}{4} spectra
are shown in Figure~\ref{fig:sim2}. In Figure~\ref{fig:sim2}a, from t=0 s to t=4 s, the
current sheet is in a slow phase when only spectra with very weak
intensity are produced by the current sheet. Typical line
profiles (Fig.~\ref{fig:sim2}b) during this phase show a double-peak shape, with
blue- and red-shifted components at about 200 km/s, corresponding to a
bi-directional jet at Alfv\'en velocity (Figure~\ref{fig:sim}a). Starting from
t=4 s, plasmoids start to appear in the current sheet, and as a
consequence, the core component of \ion{Si}{4} spectra starts growing, as
shown by the line profile at t=5.4 s. As smaller plasmoids grow and
merge with each other to form larger plasmoids, the core component
grows stronger, leading to triangular shaped line profile with a
strong core and two broad wings at t=13 s. The temporal evolution of
synthetic \ion{Si}{4} spectra from a simulated current sheet agrees well
with IRIS observations (Figure~\ref{fig:obs1}), suggesting evolving line profiles
of the UV bursts presented in this paper are driven by the
growth of the plasmoid instability.

Fig.~\ref{fig:sim3} shows histograms of the amount of plasma in different velocity intervals, which will contribute to
emission of the \ion{Si}{4} line if the local temperature is close
          to the formation temperature of \ion{Si}{4} (80,000K). Panel (a)
          shows that during the slow phase, the current sheet is
          dominated by Alfvénic flows in a bi-directional jet,
          resulting in a double-peak histogram. The bi-directional jet
          weakens when plasmoids start to appear in the current
          sheet. As the plasmoids grow, they cause fluctuations in the
          density profile of the current sheet. During the fast phase
          (panel (b)), a substantial number of plasmoids
          are formed, which move along the current sheet at highly
          fluctuating velocities. In other words, a substantial amount
          of plasmoids can move at velocities significantly lower than
          the Alfvén speed. This result agrees with the numerical
          experiments reported in \citet{37}, which suggests that plasmoids
          in the current sheet assume a broad distribution in size and
          lead to a similar distribution of velocities. The high-speed
          part of the velocity distribution is the reason for the
          Doppler-shifted wings on both blue and red sides of the line
          profiles, which are sometimes mentioned as high-velocity
          tails of the UV burst spectra, while the low-speed
          core of the velocity distribution leads to a strong central
          component in the \ion{Si}{4} spectra.          

\section{Conclusions and discussion}\label{sec:con}

In this paper, we present rare spectroscopic observations of the
transition of a reconnecting current sheet from a slow
quasi-continuous phase to a fast and impulsive phase. We find that the
observations of events showing bidirectional flows followed quickly by
triangular shaped-profiles are compatible with the growth of the
plasmoid instability, which is an important prediction of theory
but had not yet been validated by experimental or
observational evidence until now. While theoretical work on this
mechanism has been extensive, observational studies are typically more
limited because of practical considerations (e.g., limitations of
remote sensing). Thanks to the high temporal resolution and high
sensitivity of the IRIS instrument, we can now, for the first time,
shed light on the events preceding the impulsive phase of transition
region UV bursts or explosive events. The new observations presented here have the
special feature of capturing both the slow and the impulsive phases of
the events, and the close correspondence with the simulation results
make a compelling case for the interpretation based on the nonlinear
plasma instability. These observations, which go further than
previous reports \citep[e.g.,][]{22, 26, 27} on plasmoids on the Sun,
demonstrate that IRIS spectroscopic data can be used for diagnosing
rapid reconnection dynamics on the Sun, thereby opening up
opportunities for future studies that constrain theoretical models of
magnetic reconnection.

\acknowledgements{\longacknowledgment} 

\bibliographystyle{aasjournal}


\end{document}